\documentclass[twocolumn,english,showpacs,prb,floatfix]{revtex4}%
\usepackage{amsfonts}
\usepackage{amsmath}
\usepackage{amssymb}
\usepackage{graphicx}%
\begin{document}
\title{Potential super-hard Osmium di-nitride with fluorite structure:
First-principles calculations }
\author{Chang-Zeng Fan, Song-Yan Zeng}
\affiliation{Department of Material Science and Engineering, Harbin Institute of
Technology, Harbin 150001, China}
\author{Li-Xin Li, Ri-Ping Liu, Wen-Kui Wang}
\affiliation{Key Laboratory of Metastable Material Science and Technology, Yanshan
University, Qinhuangdao 066004, China}
\author{Ping Zhang }
\affiliation{Institute of Applied Physics and Computational Mathematics, Beijing 100088, China}
\author{Yu-Gui Yao}
\affiliation{Beijing National Laboratory for Condensed Matter Physics, Institute of
Physics, Chinese Academy of Sciences, Beijing 100080, China}

\begin{abstract}
We have performed systematic first-principles calculations on di-carbide,
-nitride, -oxide and -boride of platinum and osmium with the fluorite
structure. It is found that only PtN$_{2}$, OsN$_{2}$ and OsO$_{2}$ are
mechanically stable. In particular OsN$_{2}$ has the highest bulk modulus of
360.7 GPa. Both the band structure and density of states show that the new
phase of OsN$_{2}$ is metallic. The high bulk modulus is owing to the strong
covalent bonding between Os 5\textit{d} and N 2\textit{p} states and the dense
packed fluorite structure.

\end{abstract}
\maketitle

The search for novel hard materials compared to or even harder than diamond,
which has the highest measured hardness of 96 GPa\cite{Tet} and bulk modulus
of 443GPa,\cite{Kit} has a long history and has stimulated a variety of great
achievements in high-pressure research.\cite{Liu,And,Bra,Mc} Consequently,
many new superhard materials have been prepared by high-pressure technique,
especially after the laser-heated diamond-anvil cells (DACs) was invented. In
general two groups of materials are powerful candidates for super-hard
materials: (i) strong covalent compounds formed by light elements, such as
polymorphy of C$_{3}$N$_{4,}$\cite{Rig} B$_{6}$O,\cite{He} and $c$-BC$_{2}%
$N.\cite{Pan} (ii) Partially covalent heavy transition metal boride, carbide,
nitride and oxide. RuO$_{2}$\cite{Ben} and OsB$_{2}$\cite{Cum} are such
examples. Theoretically, the nature of hardness has been extensively
investigated and many new models have been
proposed.\cite{Tet,Liu,Petrescu,Gao,HeJL,Cezh,Gao1} For the strong covalent
materials, hardness can be directly derived,\cite{Gao,HeJL,Cezh,Gao1} while
for some metallic transition metal-based super-hard materials, it is
acknowledged that bulk modulus or shear modulus can measure the hardness in an
indirect way.\cite{Tet,Petrescu,Mattesini} That is, materials with high bulk
or shear modulus are likely to be hard materials.

In the present paper, we focused on the bulk modulus, mechanical and
energetic stability of osmium di-nitride (OsN$_{2}$) with CaF$_{2}$
structure\cite{Pra} by calculating the elastic constants within the
density functional based electronic structure method.\cite{Hoh} For
the first time we report that the proposed OsN$_{2}$ compound has
very high value of bulk modulus (360.7 GPa) which is even higher
than that of OsO$_{2}$ with the same structure (347.5 GPa) and is
comparable with that of orthorhombic OsB$_{2}$ (365-395
GPa\cite{Cum}).

All first-principles calculations were performed with the CASTEP
code.\cite{Seg} The ultrasoft pseudopotential (USPP)\cite{Van} was employed to
describe the interaction between ions and electrons. Both the local-density
approximation (LDA)\cite{Cep} and the generalized gradient approximation
(GGA)\cite{Per} were used to describe the exchange and correlation potentials.
For the Brillouin-zone sampling, the Monkhorst-Pack (MP) scheme with a grid of
0.03 \AA $^{-1}$ was adopted.\cite{Mon} The plane-wave cutoff energy is chosen
to be 550 eV for LDA and 500 eV for GGA calculations. For the self-consistent
field iterations, the convergence was assumed when (i) the total energy
difference between the last two cycles was less than 1$\times$10$^{-6}$
eV/atom; (ii) the maximal force on each atom was below 0.006 eV \AA $^{-1}$,
and (iii) the maximal atomic displacement was below 2$\times1$0$^{-4}$ \AA .
We have tested that with even more strict parameters the total energy can be
converged within 0.002 eV/atom for all the systems studied. After getting the
equilibrium geometry configuration, we applied the so-called \textquotedblleft
stress-strain\textquotedblright\ method to obtain the elastic constants in
that the stress can be easily obtained within the density functional based
electronic structure method.\cite{Nielsen} The \textquotedblleft
stress-strain\textquotedblright\ relation can be described as%
\begin{table*}
\caption{The calculated equilibrium lattice parameters $a$(\AA ),
elastic constants $c_{ij}$ (GPa), bulk modulus $B$ (GPa),
polycrystalline shear modulus $G$ (GPa), Young's modulus $E$ (GPa),
and Poisson's ratio $\nu$ of
typical pure crystals}%
\label{table1}%
\begin{ruledtabular}
\begin{tabular}
[c]{ccccccccccccc} &  & $a$ & $c_{11}$ & $c_{33}$ & $c_{44}$ &
$c_{13}$ & $c_{12}$ & $B$ & $G$ & $E$ & $\nu$ & \\\hline Diamond &
LDA & 3.525 & 1105.8 &  & 607.3 &  & 140.5 & 462.3 & 545.0 &
1173.8 & 0.08 & \\
& GGA & 3.566 & 1053.3 &  & 569.1 &  & 119.5 & 430.7 & 518.0 &
1109.3 & 0.07 &
\\
& Ave. & 3.546 & 1079.6 &  & 588.2 &  & 130.0 & \textbf{446.5} &
531.5 &
1141.6 & 0.08 & \\
& Expt. & 3.567$^{a}$ &  &  &  &  &  & 443$^{b}$ &  &  &  & \\
Pt (fcc) & LDA & 3.921(3.890$^{c}$) & 391.1 &  & 82.3 &  & 279.0 &
316.4(320$^{c}$) & 69.2 & 193.5 & 0.40 & \\
& GGA & 3.998(3.967$^{c}$) & 307.9 &  & 65.7 &  & 232.9 & 257.9(
238$^{c}$) &
51.6 & 145.1 & 0.41 & \\
& Ave. & 3.960(3.928$^{c}$) & 349.5 &  & 74.0 &  & 256.0 & \textbf{287.2}%
(279$^{c}$) & 60.4 & 169.3 & 0.41 & \\
& Expt. & 3.924$^{d}$ &  &  &  &  &  & 276$^{e}$ &  &  &  & \\
Os (fcc) & LDA & 3.798(3.814$^{f}$) & 686.9 &  & 361.3 &  & 323.1 &
444.4(441.3$^{f}$) & 271.6 & 676.9 & 0.25 & \\
& GGA & 3.851(3.851$^{f}$) & 614.7 &  & 328.0 &  & 282.5 & 393.2(392.9$^{f}%
$) & 247.1 & 612.9 & 0.24 & \\
& Ave. & 3.824(3.841$^{f}$) & 650.8 &  & 344.7 &  & 302.8 & \textbf{418.8}%
(417.1$^{f}$) & 259.4 & 644.9 & 0.25 & \\
Os (hcp) & LDA & 2.712 & 808.7 & 888.6 & 271.2 & 264.7 & 243.7 &
449.0 &  &  &
& \\
& GGA & 2.750(2.746$^{a}$) & 730.1 & 798.3 & 246.9 & 230.5 & 209.8 &
398.9 &
&  &  & \\
& Ave. & 2.731 & 1538.8 & 843.5 & 259.1 & 247.6 & 226.8 &
\textbf{424.0 } &  &
&  & \\
& Expt. & 2.7313$^{g}$ &  &  &  &  &  & 395$^{g}$,462$^{h}$ &  &  &
&
\\ 
\end{tabular}
\end{ruledtabular}
$^{a}$Reference 12.                 $^{b}$Reference 2.
$^{c}$Reference 26, 37.             $^{d}$Reference \cite{Pearson}.
$^{e}$Reference \cite{Brandes}.     $^{f}$Reference 28.
$^{g}$Reference 33.                 $^{h}$Reference 29.
\end{table*}

\begin{equation}
(\sigma_{1},\sigma_{2},\sigma_{3},\sigma_{4},\sigma_{5},\sigma_{6}%
)=C(\varepsilon_{1},\varepsilon_{2},\varepsilon_{3},\varepsilon_{4}%
,\varepsilon_{5},\varepsilon_{6})^{T}. \tag{1}\label{e1}%
\end{equation}
For the cubic crystal, there are only three non-zero independent symmetry
elements (\textit{c}$_{11}$, \textit{c}$_{12}$ and \textit{c}$_{44}$). Appling
two kinds of strains ($\varepsilon_{1}$ and $\varepsilon_{4}$) along the
crystallographic directions shown in Fig. 1(a)-(b), respectively, can give
stresses relating to these three elastic coefficients, yielding an efficient
method for obtaining elastic constants for the cubic system. For the hexagonal
crystal, there are five independent symmetry elements (\textit{c}$_{11}$,
\textit{c}$_{12}$, \textit{c}$_{13}$, \textit{c}$_{33}$ and \textit{c}$_{44}%
$). In order to obtain the additional components \textit{c}$_{13}$ and
\textit{c}$_{33}$, another strain [$\varepsilon_{3}$, see Fig. 1(c)] is
needed. For each strain, in our practical calculations, its value is varied
from -0.003 to +0.003 with a step of 0.0012, then each of three elastic
constants takes the arithmetic average value of the six steps. The bulk
modulus is obtained from the elastic constants by the relation \textit{B}%
=(\textit{c}$_{11}$+2\textit{c}$_{12}$)/3.

\begin{figure}[ptb]
\begin{center}
\includegraphics[width=1.0\linewidth]{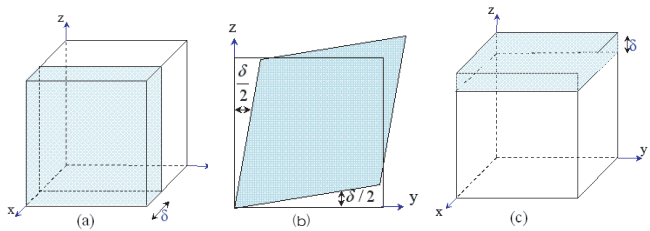}
\end{center}
\caption{(Color online) The schematic of strain types: (a)$\varepsilon_{1}%
$;(b)$\varepsilon_{4}$; (c)$\varepsilon_{3}$}%
\label{fig1}%
\end{figure}

The lattice and elastic constants of diamond, pure platinum,
hexagonal and cubic pure osmium were calculated to verify the
reliability of the present calculations. It is well known that LDA
usually underestimates the lattice constants and overestimates the
elastic constants, while GGA overestimates the lattice constants and
underestimates the elastic constants.\cite{Yu,Stam} For this reason
we adopted to use the average of the LDA and GGA results as our
theoretical estimates. As shown in Table I, the theoretical average
lattice constant and bulk modulus of diamond are 3.546 \AA and 446.5
GPa, which agree well with the experimental values of 3.567 \AA
\cite{Petrescu} and 443 GPa,\cite{Kit} with an error of 0.59\% and
0.79\%, respectively. For the cubic osmium and cubic platinum, the
theoretical bulk modulus values (Pt: 287.2 GPa, Os: 418.8 GPa) are
also in accordance with other theoretical results (Pt: 279
GPa,\cite{Yu} Os: 417.1 GPa\cite{Sa}). With regard to the hexagonal
Os, as shown in Table I, the calculated equilibrium lattice
parameter $c/a$ is 1.581, in good agreement with the experimental
value of 1.580.\cite{Cyn,Fas} The calculated bulk modulus for
hexagonal Os is 424.0 GPa, which is also in accordance with previous
theoretical results (403 GPa,\cite{Zhe} 429.2 GPa\cite{Sa}) and
experimental measurements (462$\pm1$2 GPa,\cite{Cyn} 411$\pm6$
GPa,\cite{Occ} 395$\pm1$5 GPa\cite{Ken}). Based on above-mentioned
accordance, therefore, we believe that the plane-wave ultrasoft
pseudopotential (PW-PP) method we employed is reliable in
investigating the mechanical properties of osmium and platinum
compounds.

Now we turn to fully study OsN$_{2}$ with fluorite structure. The
results of lattice constant, elastic constants, and bulk modulus of
OsN$_{2}$ are listed in Table II. For comparison, we have in
addition given a calculation on platinum dinitride (PtN$_{2}$) and
osmium dioxide (OsO$_{2}$), and the results are also listed in Table
II. Given the elastic constants and bulk modulus, the shear modulus
$G$ and the Young's modulus \textit{E} can be deduced as
follows: $G=$($c_{11}-c_{12}+2c_{44}$)/4, $E=9BG/(3B+G)$ and \textit{v}%
=$E/(2G)-1.$ These quantities are also shown in Table II.
\begin{table*}
\caption{The calculated equilibrium lattice parameters $a$(\AA ),
elastic constants $c_{ij}$ (GPa), bulk modulus $B$ (GPa),
polycrystalline shear modulus $G$ (GPa), Young's modulus $E$ (GPa),
and Poisson's ratio $\nu$ of
some fluorite and pyrite crystals}%
\label{table2}%
\begin{ruledtabular}
\begin{tabular}
[c]{ccccccccccccc} &  & $a$ & $c_{11}$ &  & $c_{44}$ & & $c_{12}$ &
$B$ & $G$ & $E$ & $\nu$ &
\\\hline
OsO$_{2}$ (fluorite) & LDA & 4.770(4.763$^{a}$) & 721.3 &  & 243.1 &
&
206.6 & 378.2(411$^{a}$,392$^{b}$) & 250.2 & 615.0 & 0.23 & \\
& GGA & 4.861 & 632.2 &  & 211.2 &  & 158.9 & 316.7 & 223.9 & 543.6
& 0.21 &
\\
& Ave. & 4.816 & 676.8 &  & 227.0 &  & 182.8 & \textbf{347.5} &
237.1 &
579.3 & 0.22 & \\
PtN$_{2}$(fluorite) & LDA & 4.943(4.866$^{c}$) & 499.9 &  & 87.4 & &
232.6 &
321.7(316$^{c}$) & 110.5 & 297.4 & 0.35 & \\
& GGA & 5.040(4.958$^{c}$) & 427.9 &  & 77.5 &  & 188.6 &
268.3(264$^{c}$) &
98.6 & 263.5 & 0.34 & \\
& Ave. & 4.992(4.912$^{c}$) & 463.9 &  & 82.5 &  & 210.6 &
\textbf{295.0
}(290$^{c}$) & 104.6 & 280.5 & 0.35 & \\
PtN$_{2}$(pyrite) & GGA & 4.874 (4.875$^{d}$) & 689 &  & 129 &  &
102 & 297.8
(278$^{d}$) & 211.3 & 512.7 & 0.21 & \\
& GGA$^{e}$ & 4.862 & 668 &  & 99 &  & 167 & 272 & 184 & 452 & 0.23 & \\
OsN$_{2}$(fluorite) & LDA & 4.781 & 544.5 &  & 103.9 &  & 309.8 &
388.0 &
117.4 & 319.9 & 0.36 & \\
& GGA & 4.856 & 465.4 &  & 79.7 &  & 267.3 & 333.3 & 89.4 & 246.2 & 0.38 & \\
& Ave. & 4.819 & 505.0 &  & 91.8 &  & 288.6 & \textbf{360.7} & 103.4
& 283.1 &
0.37 & \\
OsN$_{2}$(pyrite) & GGA & 4.925 & 523 &  & 107 &  & 213 & 316 & 131
& 345.3 & 0.32 & \\
\end{tabular}
\end{ruledtabular}
$^{a}$Reference 35. $^{b}$Reference 36. $^{c}$Reference 26, 37.
$^{d}$Reference 42. $^{e}$Reference 41.
\end{table*}

The key criteria for mechanical stability of a crystal is that the strain
energy must be positive,\cite{Nye} which means in a hexagonal crystal that the
elastic constants should satisfy the following inequalities,%

\begin{equation}
c_{44}>0,c_{11}>|c_{12}|,(c_{11}+c_{12})c_{33}>2c_{13}^{2}, \tag{2}\label{e2}%
\end{equation}
while for a cubic crystal,%

\begin{equation}
c_{44}>0,c_{11}>|c_{12}|,c_{11}+2c_{12}>0.\tag{3}\label{e3}%
\end{equation}
It is straightforward to verify from Table I that the elastic
constants of the hexagonal osmium satisfy formula (\ref{e2}),
implying the stability of hcp Os, which is consistent with the
experimental observation. In the same manner, from our calculation
results in Table I, one can find that PtN$_{2}$, OsN$_{2}$ and
OsO$_{2}$ with fluorite structure are also mechanically stable since
their elastic constants fit well in formula (\ref{e3}). The
stability of these three crystals can also be confirmed by providing
the Possion's ratio, whose value is usually between -1 and 0.5,
corresponding to the lower and upper limit where the materials do
not change their shapes. Note that the present result of bulk
modulus of PtN$_{2}$ is 295.2 GPa. The previous FLAPW
(full-potential linearized augmented plane waves) calculation gives
290 GPa.\cite{Yu} Remarkably, the two approaches agree well,
suggesting again the reliability of PW-PP method in exploring the
structural properties of transition metal compounds. On the other
hand, we obtained the bulk modulus of OsO$_{2}$ to be 347.5 GPa,
which is $\sim$13\% smaller than that obtained from full-potential
linear muffin-tin orbital (FP-LMTO) method.\cite{Lun,Hug} This
difference may come from different density functional based
electronic structure method. It reveals in Table I that OsN$_{2}$
has the highest bulk modulus of 360.7 GPa in our series of
calculations, this value of OsN$_{2}$ is much higher than that of
other noble metal di-nitride (347 GPa for IrN$_{2}$, 190 GPa for
AgN$_{2}$, and 222 GPa for AuN$_{2}$\cite{Yu2}). The shear modulus
of OsN$_{2}$ is calculated to be 103.4 GPa, comparable with that of
PtN$_{2}$ (104.6 GPa) as shown in Table I, but much smaller than
those of diamond (531.5 GPa) and OsO$_{2}$ (237.1 GPa). Thus
compared to diamond or OsO$_{2}$, OsN$_{2}$ cannot withstand the
shear stress to a large extent. It is interesting to point out that
OsN$_{2}$ was implicitly referred in Ref.\cite{Yu2} to be unstable
or has a little bulk modulus. This inconsistency may come from the
intrinsic technique flaws in WIEN2K code for obtaining the elastic
constants, in which the implanted rhombohedral distortion is not
volume-conservative. When applying a
non-volume-conservative strain, the expression $\phi_{elast}=\frac{V_{0}}%
{2}c_{ij}\varepsilon_{i}\varepsilon_{j}$ for elastic energy is no longer
accurate, resulting in unexpected results in some special cases\cite{Yu4}.

\begin{figure}[ptb]
\begin{center}
\includegraphics[width=1.0\linewidth]{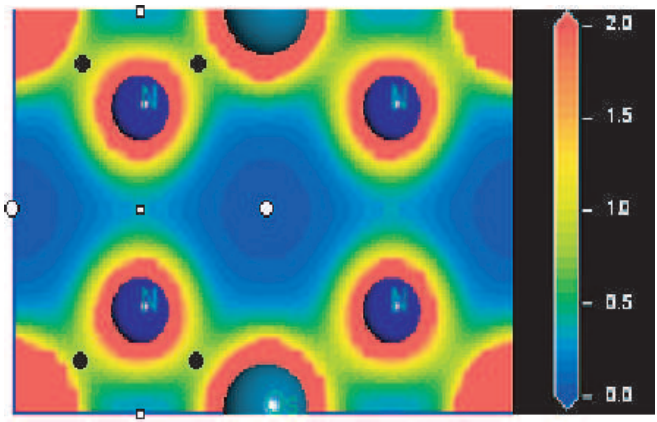}
\end{center}
\caption{(Color online) Total electron density of OsN$_{2}$ at the
($\bar {1}10$) plane. The density at three types symmetry points
(they are labeled with filled circles, open squares and open
circles) are approximately 0.8, 0.3
and 0 $e$ \AA $^{-3}$.}%
\label{fig2}%
\end{figure}

Furthermore, the electronic structure and chemical bonding of
OsN$_{2}$ with fluorite structure are studied by calculating its
total charge density, Mulliken population, and density of state
(DOS). In Fig. 2, we plot the total electron density in a (\={1}10)
plane which cut through both the Os and N sites. The bonding
behavior of OsN$_{2}$ can be effectively revealed by analyzing the
charge density data in real space $\rho(r)$ at three types
crystalline symmetry points, as indicated by filled circles, open
squares, and open circles in Fig. 2. We found that the charge
density of these three kinds points are about 0.8, 0.3 and 0
\textit{e} \AA $^{-3}$ respectively. Thus the charge density maximum
lies between Os and N atoms, indicating formation of strong covalent
bonding between them. Combining the fact that each N atom occupies
the tetrahedral interstitial formed by four Os atoms around it, it
is not difficult to understand that OsN$_{2}$ has a low
compressibility.
\begin{table}[th]
\caption{The calculated Atomic and bond Mulliken population analysis
of OsN$_{2}$, OsO$_{2}$ and PtN$_{2}$. NM1 and NM2 denote the first
and second
non-metal atoms, and M denotes the metal atom.}%
\label{table3}%
\begin{tabular}
[c]{cccccccc}\hline\hline &  &  & {Atomic (e)} &  &  & Bond (e) &
\\\hline &  & NM1 & NM2 & M & NM1-M & NM2-M & NM1-NM2\\\hline
OsN$_{2}$ &  & -0.54 & -0.54 & 1.09 & 1.25 & 1.25 & -0.7\\
OsO$_{2}$ &  & -0.55 & -0.55 & 1.10 & 0.98 & 0.98 & -0.45\\
PtN$_{2}$ &  & -0.45 & -0.45 & 0.9 & 1.08 & 1.08 &
-0.34\\\hline\hline
\end{tabular}
\end{table}
Table III shows bond Mulliken population analysis of OsN$_{2}$,
OsO$_{2}$, and PtN$_{2}$. It indicates that for these three kind of
materials the bonding is formed between metal atom and non-metal
atom, while a weak-bonding is formed between two non-metal atoms.
This is compatible to analysis of the electron density of OsN$_{2}$
in Fig. 2. Table III also lists the Mulliken atomic population
analysis results, from which we can see the total charge transfer
from Os to N is 1.10, resulting Os in +1.10 charge state and N
--0.55 charge state. Therefore, the chemical bonding between Os and
N has some character of ionicity. It shows in Table III that the
transferred charge in OsN$_{2}$ is almost the same as that in
OsO$_{2}$, and is more than that in PtN$_{2}$. Thus we can make a
conclusion that the charge transfer effect is more influenced by the
metal atom rather than the non-metal atom. It is interesting to note
that the mechanical properties of OsN$_{2}$ are also very similar to
OsO$_{2}$ rather than PtN$_{2}$, as revealed in our above
discussions.

\begin{figure}[ptb]
\begin{center}
\includegraphics[width=1.0\linewidth]{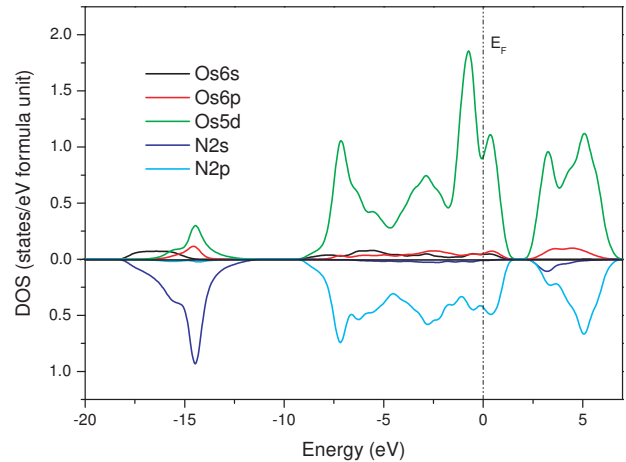}
\end{center}
\caption{(Color online) Partial densities of states of OsN$_{2}$.}%
\label{fig3}%
\end{figure}

The partial DOS is shown in Fig. 3, no energy gap near Fermi level
is seen, indicating the metallic nature of OsN2. At the Fermi level
the total DOS is 1.88 states/eV formula units. It reveals that from
-18 to -12 eV the states are mainly N (2$s$) states with a small
contribution from Os (5$d$ and 6$p$). The states above -9 eV mainly
come from Os 5$d$ and N 2$p$ orbitals.

It was recently demonstrated in both theory and experiment that the
synthesized platinum nitride crystallized in pyrite
structure.\cite{Yu3,Crow,Young} The pyrite structure (space group number 205),
which was also observed in the silica recently,\cite{Kuwa} is cubic with 12
atoms per primitive cell. For pyrite PtN$_{2}$, both the four Pt atoms and the
midpoints of the four nitrogen pairs arrange in fcc positions and result in
NaCl-type arrangement. In addition, each pair of nitrogen atoms aligns along
one of the (111) directions. Besides the lattice constant $a$, the position
(\textit{u}, \textit{u}, \textit{u}) of N is the only free structural
parameter of pyrite PtN$_{2}$. Inspired by these advances, we have also
performed a series of \textit{ab initio} total energy calculations to find if
OsN$_{2}$ favors the intriguing pyrite structure as platinum nitride does.
Figure 4(a) gives the the energy of OsN$_{2}$ with the internal parameter
\textit{u} varied from 0.23 to 0.40. The plot reveals that fluorite OsN$_{2}%
$\ lies at a local minimum, indicating its metastable nature. The location of
lowest total energy is at 0.3614 for GGA calculations, corresponding to the
lattice constant of 4.9246 \r{A} of the pyrite structure. The bond length of
nitrogen pairs in pyrite OsN$_{2}$\ is found to be 1.365\ \r{A}, even smaller
than that of pyrite PtN$_{2}$ (1.51 \r{A}). The separation of nitrogen pairs
in the pyrite OsN$_{2}$\ and PtN$_{2}$ is nearly the same as that of a
single-bonded cubic-gauche structure of N observed
recently,\cite{Eremets,Peng} which means that the nitrogen pairs probably
consist some characteristics of covalent bonding. The calculated formation
energies of pyrite and fluorite OsN$_{2}$ at ambient pressure are 0.69 eV and
1.10 eV (per formula unit), which are smaller than those of pyrite PtN$_{2}$
(0.72eV) and fluorite PtN$_{2}$ (3.5eV) respectively. Figure 4(b) plots the
\begin{figure}[ptb]
\begin{center}
\includegraphics[width=1.0\linewidth]{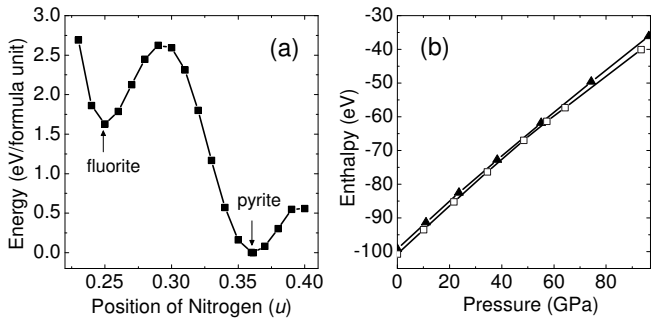}
\end{center}
\caption{(a) Total energy of OsN$_{2}$ versus the internal parameter
$u$; (b) Total energy as a function of the volume per formula unit
for OsN$_{2}$ with
fluorite and pyrite structure.}%
\label{fig4}%
\end{figure}
enthalpy versus pressure for the fluorite and pyrite structures. In
the overall range of external pressure that we have studied, the
enthalpy of pyrite OsN$_{2}$\ is always lower than that of fluorite
OsN$_{2}$, implying that no first-order phase transition might occur
at zero-temperature between
these two structures. In addition, the elastic constants of pyrite OsN$_{2}%
$\ and PtN$_{2}$\ are also calculated and listed in Table II, wherein the
results for PtN$_{2}$ show good agreement with those of experimental and other
available theoretical results. The calculated elastic constants of pyrite
OsN$_{2}$ satisfy formula (\ref{e3}). Therefore, it is also a mechanically
stable crystal structure though its bulk modulus is about 10\% lower than that
of fluorite OsN$_{2}$.

In conclusion, the OsN$_{2}$ with fluorite structure is first reported to be
mechanically stable and have a very high bulk modulus of 360.7 GPa by the
first-principles calculations. The electronic and chemical bonding properties
have been investigated, indicating that the bonding is a mixture of covalent
and ionic components. It is found that the electronic properties of OsN$_{2}$
is very similar to that of PtN$_{2}$ with the same structure. As a pyrite-type
PtN$_{2}$\ and a orthorhombic-type OsN$_{2}$ have been very recently
synthesized under high pressure and high temperature
conditions,\cite{Mao,Crow,Young2} we expect that the OsN$_{2}$ as well as
PtN$_{2}$ with fluorite structure may be experimentally prepared in the future.

We are also grateful to Dr. R. Yu for useful discussions and the two
anonymous referees for fruitful suggestions. This work is supported
by NSFC Nos. 10404035, 10534030, 50325103, and 10544004.


\begin{thebibliography}{99}                                                                                               %


\bibitem {Tet}D. M. Teter, MRS Bull. \textbf{23}, 22 (1998).

\bibitem {Kit}C. Kittel, \textit{Introduction to Solid State Physics}, 8th
edition (Wiley, New York, 1996).

\bibitem {Liu}A. Y. Liu, and M. L. Cohen, Science \textbf{245}, 841 (1989).

\bibitem {And}R. A. Andrievski, Inter. J. Refract. Meta. \& Hard Mater.
\textbf{19}, 447 (2001).

\bibitem {Bra}V. V. Brazhkin, A. G. Lyapin, and R. J. Hemley, Phil. Mag. A
\textbf{82}, 231 (2002).

\bibitem {Mc}P. F. Mcmillan, Nat. Mater. \textbf{1}, 19 (2002).

\bibitem {Rig}G. -M. Rignanese, J. -C. Charlier, and X. Gonze, Phys. Rev. B
\textbf{66}, 205416 (2002).

\bibitem {He}D. W. He, Y. S. Zhao, L. Daemen, J. Qian, T. D. Shen, and T. W.
Zerda, Appl. Phys. Lett. \textbf{82}, 643 (2002).

\bibitem {Pan}Z. C. Pan, H. Sun, and C. F. Chen, Phys. Rev. B \textbf{70},
174115 (2004).

\bibitem {Ben}K. Benyahia, Z. Nabi, A. Tadjer, and A. Khalfi, Physica. B
\textbf{339}, 1 (2003).

\bibitem {Cum}R. W. Cumberland, M. B. Weinberger, J. J. Gilman, S. M. Clark,
S. H. Tolbert, and R. B. Caner, J. Am. Chem. Soc. \textbf{127}, 7264 (2005).

\bibitem {Petrescu}M. I. Petrescu, Diam. Relat. Mater. \textbf{13,} 1848 (2004).

\bibitem {Gao}F. M. Gao, J. L. He, E. D. Wu, S. M. Liu, D. L. Yu, D. C. Li, S.
Y. Zhang, and Y. J. Tian, Phys. Rev. Lett. \textbf{91}, 015502 (2003).

\bibitem {HeJL}J. L. He, E. D. Wu, H. T. Wang, R. P. Liu, and Y. J. Tian,
Phys. Rev. Lett. \textbf{94}, 015504 (2005).

\bibitem {Cezh}A. \v{S}im\r{u}nek, and J. Vack\'{a}\v{r}, Phys. Rev. Lett.
\textbf{96}, 085501 (2006).

\bibitem {Gao1}F. M. Gao, Phys. Rev. B \textbf{73}, 132104 (2006).

\bibitem {Mattesini}M. Mattesini, R. Ahuja, and B. Johansson, Phys. Rev. B
\textbf{68}, 184108 (2003).

\bibitem {Pra}Practically, systematic calculations were performed on
di-carbide, -nitride, -oxide and -boride of platinum and osmium with the
fluorite structure respectively, but only found that PtN$_{2}$, OsN$_{2}$ and
OsO$_{2}$ are mechanically stable. The unstable phases are not shown here.

\bibitem {Hoh}P. Hohenberg, W. Kohn, Phys.Rev. \textbf{136}, B864 (1964); W.
Kohn, L. J. Sham, Phys. Rev. \textbf{140}, A1133 (1965).

\bibitem {Seg}M. D. Segall, P. L. D. Linda, M. J. Probert, C. J. Pickard, P.
J. Hasnip, S. J. Clark, and M. C. Payne, J. Phys.: Condens. Matter
\textbf{14}, 2717 (2002).

\bibitem {Van}D. Vanderbilt, Phys. Rev. B \textbf{41}, 7892 (1990). For the
GGA calculations, the pseudopotential generated by using the related Perdew-
Burke-Ernzerhof type of exchange-correlation functions should be used.

\bibitem {Cep}D. M. Ceperley, and B. J. Alder, Phys. Rev. Lett. \textbf{45},
566 (1980).

\bibitem {Per}J. P. Perdew, K. Burke, and W. Ernzerhof, Phys. Rev. Lett.
\textbf{77}, 3865 (1996).

\bibitem {Mon}H. J. Monkhorst, and J. D. Pack, Phys. Rev. B \textbf{13}, 5188 (1976).

\bibitem {Nielsen}O. H. Nielsen, and R. M. Martin,Phys. Rev. B \textbf{32},
3792 (1985).

\bibitem {Yu}R. Yu, and X. F. Zhang, Appl. Phys. Lett. \textbf{86}, 121913 (2005).

\bibitem {Stam}C. Stampfl, W. Mannstadt, R. Asahi, and A. J. Freeman, Phys.
Rev. B \textbf{63}, 155106 (2001).

\bibitem {Sa}B. R. Sahu, and Leonard Kleinman, Phys. Rev. B \textbf{72},
113106 (2005).

\bibitem {Cyn}H. Cynn, J. E. Klepeis, C. S. Yoo, and D. A. Young, Phys. Rev.
Lett. \textbf{88}, 135701 (2002).

\bibitem {Fas}L. Fast, J. M. Wills, B. Johansson, and O. Eriksson, Phys. Rev.
B \textbf{51}, 17431 (1995).

\bibitem {Zhe}J. C. Zheng, Phys. Rev. B \textbf{72}, 052105 (2005).

\bibitem {Occ}F. Occelli, D. L. Farber, J. Badro, C. M. Aracne, D. M. Teter,
M. Hanfland, B. Canny, and B.Couzinet, Phys. Rev. Lett. \textbf{93}, 095502 (2004).

\bibitem {Ken}T. Kenichi, Phys. Rev. B \textbf{70}, 012101 (2004).

\bibitem {Nye}J. F. Nye, \textit{Physical Properties of Crystals} (Oxford
University Press, Oxford, 1985).

\bibitem {Lun}U. Lundin, L. Fast, L. Nordstrom, B. Johansson, J. M. Wills, and
O. Eriksson, Phys. Rev. B \textbf{57}, 4979 (1998).

\bibitem {Hug}H. W. Hugosson, G. E. Grechnev, R. Ahuja, U. Helmersson, L. Sa,
and O. Eriksson, Phys. Rev. B \textbf{66}, 174111 (2002).

\bibitem {Yu2}R. Yu, and X. F. Zhang, Phys. Rev. B \textbf{72}, 054103 (2005).

\bibitem {Yu4}R. Yu, Private communication.

\bibitem {Mao}E. Gregoryanz, C. Sanloup, M. Somayazulu, J. Badro, G. Flquet,
H.K. Mao, and R. J. Hemley, Nat. Mater. \textbf{3}, 294 (2004).

\bibitem {Kuwa}Y. Kuwayama, K. Hirose, N. Sata, and Y. Ohish, Science
\textbf{309,} 923 (2005).

\bibitem {Yu3}R. Yu, Q. Zhang, and X. F. Zhang, Appl. Phys. Lett. \textbf{88},
051913 (2006).

\bibitem {Crow}J. C. Crowhurst, A. F. Goncharov, B. Sadigh, C. L. Evans, P. G.
Morrall, J. L. Ferreira, and A. J. Nelson, Science \textbf{311}, 1275 (2006).

\bibitem {Young}A. F. Young, J. A. Montoya, C. Sanloup, M. Lazzeri, E.
Gregoryanz, and S. Scandolo, Phys. Rev. B \textbf{73}, 153102 (2006).

\bibitem {Eremets}M. I. Eremets, A. G. Gavriliuk, I. A. Trojan, D. A.
Dzivenko, and R. Boehler, Nat. Mater. \textbf{l3}, 558 (2004).

\bibitem {Peng}T. Zhang, S. Zhang, Q. Chen, and L. M. Peng, Phys. Rev. B
\textbf{73}, 094105 (2006).

\bibitem {Young2}A. F. Young, C. Sanloup, E. Gregoryanz, S. Scandolo, R. J.
Hemiey, and H. K. Mao, Phys. Rev. Lett. \textbf{96} 155501 ( 2006).

\bibitem {Pearson}W. B. Pearson, \textit{A Handbook of Lattice Spacing and
Structures of Metals and Alloys}, (Pergamon, New York, 1958).

\bibitem {Brandes}E. A. Brandes, \textit{Smithells Metal Reference Book}, 6th
ed. (Butterworth, London, 1983).
\end{thebibliography}
\end{document}